\documentclass[12pt,eqsecnum]{revtex4}
\def\beq{\begin{equation}}
\def\eeq{\end{equation}}
\def\bea{\begin{eqnarray}}
\def\eea{\end{eqnarray}}

\begin{document}

\title{ Relating the generating functionals in field/antifield formulation
 through finite field dependent BRST transformation
  }


\author{ Sumit Kumar Rai \footnote{e-mail address:- sumitssc@gmail.com}}
\author{ Bhabani Prasad Mandal \footnote{e-mail address:
\ \ bhabani@bhu.ac.in, \ \ bhabani.mandal@gmail.com  }}


\affiliation{ Department of Physics,\\
Banaras Hindu University,\\
Varanasi-221005, INDIA. \\
}


\begin{abstract}


We study the field/antifield formulation of pure  Yang Mills 
theory in the framework of finite field dependent BRST transformation. We show
 that the generating functionals corresponding to different solutions
 of quantum  master equation are connected through the finite field dependent
 BRST transformations. We establish this result with the help of several explicit examples.                                                                          
\end{abstract}
\maketitle
\section{Introduction}
BRST symmetries are extremely useful in quantum field theory and play important role in the 
discussion of quantization, renormalization, unitarity and other aspects of gauge theories \cite{brs,ht,wei}. 
The nilpotent BRST transformation is characterized by an infinitesimal, anticommuting and space time independent parameter and leaves the effective action invariant. Joglekar and Mandal \cite{jm} generalized the BRST transformation where the parameter involved is 
finite and field dependent but space-time independent. Such finite field dependent 
BRST (FFBRST) transformations are also nilpotent and leave the Faddeev-Popov (FP) effective action invariant.
However, the path integral measure changes in a  non-trivial way due to the 
finite field dependent parameter in such transformations. It has been shown \cite{jm} that the non-trivial Jacobian of the path integral measure can always be expressed as $e^{iS_1}$, where $S_1$ is some local function of 
field variables and can be a part of the effective action. Thus, FFBRST can connect the 
generating functionals of two different effective field theories with suitable choice of  
finite field dependent parameter \cite{jm}.
For example, these can be used to connect FP
effective action in Lorentz gauge with a gauge parameter $\lambda $ to
(i) the most general BRST/anti-BRST symmetric action in Lorentz gauges \cite{jm}
, (ii) the FP effective action  in axial gauge \cite{sdj0}, (iii)  the 
FP effective action  in Coulomb gauge \cite{cou},
(iv) FP effective action  with another distinct gauge parameter 
$\lambda^\prime $ \cite{jm} and (v) the FP effective action in quadratic gauge \cite{jm}.      
  The choice of the parameter  is crucial in connecting different
effective gauge theories by means of the FFBRST.

FFBRST transformations have  found many applications \cite{sdj0,cou,sdj,rb,sdj1,etc} in the 
study of gauge
 theories. Correct prescription for the poles in the gauge field propagators in 
non-covariant gauges have been derived by connecting effective theories in covariant gauges to 
the theories in non-covariant
 gauges by using FFBRST \cite{sdj1,cou}. The divergent energy integrals in Coulomb gauge are 
regularized by modifying time like propagator using FFBRST \cite{cou}.

In this present work, we have extended the FFBRST formulation for the case of field/antifield 
formulation of 
pure Yang Mills (YM) theory. We have shown, the generating functionals in field/antifield 
formulation, corresponding to different solutions of quantum master equation \cite{ht,wei,bv} 
are related through FFBRST transformations. The choice of finite field dependent BRST 
parameter plays crucial role in relating the generating functionals. A particular choice of 
FFBRST parameter connects a pair of generating functionals. We consider several choice of 
FFBRST parameter to show the connection explicitly.

This paper is organised as follows. We provide brief introduction to field/antifield 
formulation in Sec. 2.1 and a brief review of the FFBRST formulation in Sec. 2.2. In Sec. 3, 
 we develop FFBRST transformation in the auxiliary field  formulation  which will be required in the later 
sections. We 
consider several choice of FFBRST parameter to establish the connection between the different generating 
functionals in field/antifield formulation in Sec. 4. In Sec. 4.1, we consider the 
connection of generating functionals for different effective theories in different gauges. 
Connection with most general BRST/anti-BRST invariant theory is 
established in Sec. 4.2.  Sec. 5 is devoted for summary and discussion.

\section{Preliminary review}
\subsection{Field/Antifield formulation}

The Lagrangian quantization of Batalin \& Vilkovisky \cite{bv}, also known as 
field/antifield formulation is considered to be one of the most powerful
 procedures of quantization of gauge theories involving BRST symmetry
\cite{ht,wei,hen1,hen2,new,new2}. The
main idea is to construct an extended action $W_\Psi(\phi , \phi^*)$ by 
introducing antifields $\phi^*$ corresponding to each  field $\phi$
with opposite statistic. Generically $\phi $  denotes all the 
fields involved in the theory. The sum of ghost number associated to a field and its
 antifield 
is equal to -1. The generating functional can be written as 
\begin{equation}
 Z= \int D{\phi}e^{i W_{\Psi}[\phi]},
\end{equation}
where
\begin{equation}
W_{\Psi}[\phi ] = W \left[\phi,\phi ^*={\frac {\partial \Psi}{\partial {\phi}}}\right].
\end{equation}
$\Psi$ is the gauge fixed fermion and has grassman parity 1 and ghost number {-1}. The
 generating functional Z does not depend on the choice of $\Psi$.
This extended quantum action satisfies certain rich mathematical
 relation called quantum master equation \cite{wei} and is  given by
\begin{equation}
\Delta e^{iW_\Psi[\phi, \phi^*]} =0  \ \mbox{ with }\ 
 \Delta\equiv \frac{\partial_r}{
\partial\phi}\frac{\partial_r}{\partial\phi^*} (-1)^{\epsilon
+1}.
\label{mq}
\end{equation}
Master equation reflects the gauge symmetry in the zeroth order of antifields
and in the  first order of antifields it reflects nilpotency of BRST transformation.
This equation  can also be written in terms of antibrackets as
\begin{equation}
\left ( W_\Psi, W_\Psi \right ) = 2i\Delta W_\Psi,
\label{ab}
\end{equation}
where the antibracket is defined as 
\begin{equation}
\left (X,Y\right ) \equiv \frac{\partial _rX}{\partial\phi}\frac{\partial_l
Y}{\partial\phi^*}-\frac{\partial _rX}{\partial\phi^*}\frac{\partial_l
Y}{\partial\phi}.
\label{ab1}
\end{equation}

Different effective actions belonging to the same theory are solutions
of master equations.  The
solutions of  master equation are unique upto (anti-) canonical transformations
which preserve the equation (\ref{ab}). For example, the solution of master equation corresponding to the extended action in axial gauge $W^A_\Psi[\phi,\phi^*]$ is related to the solution of master equation corresponding to the extended action in
 Lorentz gauge 
 $W^L_\Psi[\phi,\phi^*]$ through (anti)-canonical transformation. In this paper, we show that the generating functionals
 corresponding
to  different solutions of master equation can 
be related  through  FFBRST transformation. In particular,
  we consider field/antifield formulation of pure YM theory to show the connection by using FFBRST transformation 
which will be reviewed briefly below.\\

\subsection{Finite field dependent BRST}
Let us now briefly review the FFBRST approach \cite{jm,sdj0,sdj,sdj1}.
 FFBRST transformations are
obtained by an integration of infinitesimal  (field dependent ) BRST
transformations  \cite{jm}. In this method all the fields are functions
of some parameter, $ \kappa : 0\le \kappa \le 1$. For a generic field $ \phi (x, \kappa),\
\phi(x, \kappa =0 ) = \phi(x) $, is the initial field and
 $ \phi(x, \kappa=1) = \phi ^\prime
(x)$ is the transformed field. Then the infinitesimal field dependent BRST transformations are
defined as
\begin{equation}
\frac{ d}{d \kappa}\phi(x, \kappa ) = \delta _{BRST }\ \phi(x, \kappa )\
\Theta ^ \prime [\phi(x,\kappa )],
\label{ibr}
\end{equation}
where $\Theta ^\prime d \kappa $ is an infinitesimal field dependent parameter.
It has been shown \cite{jm} by integrating these equations from $ \kappa=0$ to $\kappa=1$
that $\phi^\prime ( x) $ are related to $\phi(x)
$ by FFBRST transformations
\begin{equation}
\phi^\prime (x) = \phi(x) + \delta _{BRST} \ \phi(x)\ \Theta [\phi(x)],
\label{fbrs}
\end{equation}
where $ \Theta [\phi(x)] $ is obtained from $\Theta ^\prime [\phi(x)] $
through the relation
\begin{equation}
\Theta [\phi(x)] = \Theta ^\prime [\phi(x)] \frac{ \exp f[\phi(x)]
-1}{f[\phi(x)]},
\label{80}
\end{equation}
and $f$ is given by $ f= \sum_i \frac{ \delta \Theta ^\prime (x)}{\delta
\phi_i(x)} \delta _{BRST}\ \phi_i(x). $
These transformations are nilpotent and symmetry of the FP effective action. 
However, 
Jacobian of the path integral measure changes the generating functional corresponding to FP effective theory to 
the generating functional for a different effective theory.
               
The meaning of these field transformations is as follows. We consider
the vacuum expectation value of a gauge invariant functional $G[A]$
in some effective theory,
\begin{equation}
<< G[A]>> \equiv \int {\cal D} \phi \ G[A] \exp(iS_{eff}[\phi]),
\label{90}
\end{equation}
where
\begin{equation}
S_{eff} = S_0 + S_{gf}+S_g. 
\label{sf}
\end{equation}
Here, $S_0$  is the pure YM action
\begin{equation}
S_0= \int d^4x \left [-\frac{1}{4}F^{\alpha\mu\nu}F_{\mu\nu}^\alpha\right ],
 \label{s0}
\end{equation} and the gauge fixing and ghost part of the effective action in Lorentz gauge are
given as
\begin{eqnarray}
S_{gf} &=& -\frac{1}{2\lambda}\int d^4x (\partial \cdot{A^\alpha})^2 \nonumber,\\
S_g &=& -\int d^4x\left [ \bar{c}^\alpha\partial^\mu
D^{\alpha\beta}_\mu c^\beta \right ].
\label{s12}
\end{eqnarray}
The covariant derivative is defined as $ D^{ab}_\mu [A] \equiv \delta ^{ab}\partial
_\mu + g f^{abc}A^c_\mu $. 
                                                                                
Now we perform the FFBRST transformation $\phi\rightarrow \phi^\prime
 $ given by Eq. (\ref{fbrs}). Then we have 
\begin{equation}
<<G[A]>>=  <<G[A^\prime ]>> = \int {\cal D} \phi ^\prime
J[\phi^\prime ] G[A^\prime ] \exp(iS^F_{eff}[\phi^\prime ]),
\label{def}
\end{equation}
on account of BRST invariance of $S_{eff}$ and the  gauge invariance of $G[A]$.
Here $J[\phi^\prime ]$ is the Jacobian associated with FFBRST transformation and is defined
as
\begin{equation}
{\cal D} \phi = {\cal D} \phi^\prime J[\phi^\prime ].
\label{jac}
\end{equation}
Note that unlike the usual infinitesimal BRST transformation the Jacobian
for FFBRST is not unity. In fact, this non-trivial Jacobian
is the source of the new results in this formulation. 
As shown in Ref. \cite{jm} for the special case $G[A]=1$, the Jacobian
$ J[\phi^\prime ]$ can always be replaced by
$ \exp(iS_1[\phi^\prime ]$),  where $S_1(\phi^\prime)$ is some {\it local}
functional of the fields and can be added to action,
\begin{equation}
S_{eff}[\phi^\prime ] +S_1[\phi^\prime ] = S_{eff}^\prime
[\phi^\prime].
\label{s1}
\end{equation}
                                                                                
Thus the FFBRST in Eq. (\ref{fbrs}) takes the generating functional 
with effective  action $S_{eff}[\phi]$ to the generating functional
corresponding to another effective action $ S_{eff}^\prime[\phi]$.
The extra part, $ S_1(\phi)$ of the new effective action $ S^\prime_{eff}[\phi]$
depends on the choice of the parameter in FFBRST transformation.
Thus, by choosing the parameter in FFBRST transformation appropriately one can
connect the generating functionals corresponding to any two effective
 gauge theories. In particular the FFBRST of Eq. (\ref{fbrs}) with
$ \Theta ^\prime [\phi(x,\kappa)] = i \int d^4y \; \bar{c} ^\alpha (y,\kappa) \left [
F^\alpha [A( y,\kappa )] - F^{\prime \alpha }[ A(y,\kappa)] \right ] $
relates the YM theory with an arbitrary gauge fixing $F[A]$ to the
YM theory with another arbitrary gauge fixing $F^\prime [A]$ \cite
{sdj}. 
 
\section{The FFBRST in auxiliary field  formulation}
To study the role of FFBRST in the field/antifield formulation, it is
convenient to use auxiliary field (B) formulation \cite{af}. In this section, we intend to generalize the FFBRST 
formulation in B field formulation. We only mention the necessary modifications of FFBRST 
formulation in presence of  auxiliary field \cite{susk}. For simplicity, we consider the case of pure 
YM theory described by the action
\begin{equation}
 S_{eff}= \int d^4x\left [-\frac{1}{4}F^{\alpha \mu \nu }F^{\alpha}_{\mu \nu } +\frac{\lambda 
}{2}(B^\alpha) ^2 - B^{\alpha}
\partial \cdot{A^\alpha }-\bar{c}^\alpha \partial^{\mu}D^{
\alpha \beta}_\mu  c^\beta \right ]. \label{seff}
\end{equation} 
 Using the method outlined in Sec. 2.2 it is straightforward to find FFBRST transformations under which the above $S_{eff}$ remains invariant.
These transformations are as follows
\begin{eqnarray}
A^{\alpha}_\mu &\rightarrow & A^\alpha_\mu + D_\mu^{\alpha\beta}c^\beta \  \Theta
(A,c,\bar{c},B),\nonumber \\
c^\alpha &\rightarrow & c^\alpha -\frac{g}{2}f^{\alpha\beta\gamma}c^\beta c^\gamma \ \Theta(A, c, \bar{c},B), \nonumber \\
\bar{c}^\alpha &\rightarrow & \bar{c}^\alpha +B^\alpha \ \Theta(A,c,\bar{c},B),\nonumber \\
B^\alpha &\rightarrow & B^\alpha.
\label{ffb}
\end{eqnarray}
However, non-trivial modification arises in the calculation of Jacobian for this FFBRST in 
auxiliary field formulation. The Jacobian can be defined as
\begin{eqnarray}
DA(x)Dc(x)D\bar{c}(x)DB(x) &=&J(x,k) DA(x,k)Dc(x,k)D\bar{c}(x,k)DB(x,k) \nonumber \\
              &=& J(k+dk)DA(k+dk)Dc(k+dk)D\bar{c}(k+dk)DB(k+dk).\nonumber\\
                                                                 \label{jac1}
\end{eqnarray}
The transformation from $\phi(k)$ to $\phi(k+dk)$ is an infinitesimal one and one has, for its Jacobian
\begin{equation}
\frac{J(k)}{J(k+dk)}=\sum_{\phi }\pm\frac{\delta{\phi(x,k+dk)}}{\delta{\phi(x,k)}},
\end{equation}
where $\sum_{\phi }$, sums over all the fields in the measure $A_\mu^\alpha,c^\alpha,\bar{c}^\alpha,B^\alpha $ and the $ \pm $ sign refers to if $\phi$ is 
a bosonic or a fermionic field.
We evaluate the right hand side as
\begin{equation}
\int d^4 x \sum_{\alpha}\left [ \sum_{\mu}\frac{\delta A_\mu^\alpha(x,k+dk)}{\delta A_
\mu^\alpha(x,k)} - \frac{\delta c^\alpha(x,k+dk)}{\delta c^\alpha(x,k)} - \frac{\delta\bar{c}^
\alpha(x,k+dk)}{\delta\bar{c}^\alpha (x,k)} + \frac{\delta B^\alpha(x,k+dk)}{\delta B^\alpha(x,k)}
 \right ],
\end{equation}
dropping those terms which do not contribute on account of the antisymmetry of structure constant. We calculate infinitesimal Jacobian change as mentioned in \cite{jm} to be
\begin{equation}
\frac{1}{J(k)}\frac{dJ(k)}{dk}= -\int d^4x \left[(\delta A^\alpha _{\mu})\frac{\delta \Theta ^
\prime}{\delta A_\mu ^\alpha}-(\delta c^{\alpha})\frac{\delta \Theta ^\prime}{\delta c^
\alpha} -\delta \bar{c}^\alpha \frac{\delta \Theta ^\prime}{\delta \bar{c}^\alpha} \right ],\label{jc1} 
\end{equation}
since, $\delta B^{\alpha}=0 $. Further, it can be shown that the Jacobian in Eq. ({\ref
{jac1}}) can be expressed as 
$e^{iS_1[\phi]}$ by following the general procedure mentioned in Ref \cite{jm}. 
 To illustrate the procedure, we consider a simple example of FFBRST in auxiliary field 
formulation. Let us take $\Theta^\prime[\phi(y,k)]=i\gamma  \int d^4y\; 
\bar{c}^\alpha(y,k)B^\alpha(y,k)$ where $\gamma$ is an arbitrary constant parameter and then
 using Eq. ({\ref{jc1}}), we obtain
\begin{equation}
\frac{1}{J(k)}\frac{dJ(k)}{dk}= i\gamma \int d^4y \;[B^\alpha(y,k)]^2.
\end{equation}
This Jacobian J(k) will be replaced by $e^{iS_1(\phi)}$ iff \cite{jm}
\begin{equation}
\int D\phi(k)exp[iS_1+iS_{eff}]\left[\frac{1}{J(k)}\frac{dJ}{dk}-i\frac{dS_1}{dk}\right] = 0.\label{es1j}
\end{equation}
We make an ansatz,
$S_1= \xi_1(k)\int d^4x \;[B^\alpha(x,k)]^2 $ where $\xi_1(k)$ is some $\kappa$-dependent arbitrary parameter satisfying the initial condition $\xi_1(k=0)=0$. Now the condition in Eq. ({\ref{es1j}}) is satisfied only when
\begin{equation}
i(\gamma-\xi_1^\prime(k))=0. \label{deff}
\end{equation}
By solving Eq. (\ref{deff}), we get $\xi_1=\gamma k $ and the extra term in the net effective action is
\begin{equation}
S_1(k=1)=\gamma\int d^4x \;[B^\alpha(x)]^2. 
\end{equation}
The effective action in Eq. (\ref{seff}) is modified as
\begin{eqnarray}
S_{eff}+S_1 &=&  \int d^4x\left [-\frac{1}{4}F^{\alpha \mu \nu }F^{\alpha}_{\mu \nu }+\frac{\lambda}{2}(B^\alpha)^2 -B^
\alpha\partial\cdot{A^\alpha}-\bar{c}^\alpha \partial^{\mu}D^{
\alpha \beta}_\mu  c^\beta +\gamma(B^\alpha)^2\right ] \nonumber\\ 
 &=&  \int d^4x\left [-\frac{1}{4}F^{\alpha \mu \nu }F^{\alpha}_{\mu \nu }+\frac{\lambda^\prime}{2}(B^\alpha)^2 -B^
\alpha\partial\cdot{A^\alpha}-\bar{c}^\alpha \partial^{
\mu}D^{\alpha \beta}_\mu  c^\beta \right ],
\end{eqnarray}
where
\begin{equation}
 \lambda^\prime= \lambda +2\gamma. 
\end{equation}
Thus, FFBRST with parameter $\Theta^\prime[\phi(y,k)]=i\gamma  \int d^4y \;
\bar{c}^\alpha (y,k)B^\alpha(y,k)$,  connects effective action with two different distinct gauge parameter $\lambda$ and $\lambda^\prime$.\\
\\We use this auxiliary field formulation of FFBRST directly in the following sections where we consider field/antifield formulation.

\section{FFBRST in field/antifield formulation}

In this section, we use FFBRST transformation in field/antifield formulation to show that the 
different generating functionals corresponding to different solution of master equation are 
related through FFBRST. We consider few explicit examples to establish this result.
\subsection{Connecting solutions of master equation in different gauges}
{\bf\small{Case I: \ Lorentz gauge to axial gauge}}\\
\\We start with the generating functional of YM theory in Lorentz gauge 
as                
\begin{equation}
Z^L = \int D\phi \exp{\left [ i S_{eff}(A,c,\bar{c},B)\right ]},\label{zl} 
\end{equation}
where $S_{eff}(A,c,\bar{c},B),$ is given by Eq. (\ref{seff}). This generating functional can be
expressed in field/antifield formulation as
\begin{equation}
 Z^L = \int [dAdcd\bar{c} dB] \exp{\left [i\int d^4x \left \{-\frac{1}{4}F^{\alpha\mu\nu}
F^\alpha_{\mu\nu}+A^{\mu\alpha *} D^{\alpha\beta}_\mu c^\beta+ c^{\alpha *}\frac{g}{2}
f^{\alpha\beta\gamma}c^\beta c^\gamma + B^\alpha \bar{c}^{\alpha *} \right \}\right ]},
\label{zla}
\end{equation}
or, compactly 
\begin{eqnarray}
 Z^L= \int D\phi \exp{\left [iS_0(\phi) 
+i \delta\Psi^L \frac{1}{\Lambda }\right ]}\equiv \int D\phi \exp{\left [iW^L_\Psi(\phi, \phi^*)\right ]},
\nonumber
\end{eqnarray}
where the gauge fixed fermion 
\begin{equation}
\Psi^L =\int d^4 x \;\bar{c}^\alpha\left [\frac{\lambda}{2} B^\alpha - \partial\cdot A^\alpha \right ].
\end{equation}
$\delta\Psi^L$  is the BRST variation of $\Psi^L$ and $\Lambda$ is the infinitesimal, anticommuting parameter of the usual BRST transformation.
 The antifields $A_\mu ^{\alpha *}, c^{\alpha *
},\bar{c}^{\alpha *}, B^{\alpha *}$ corresponding to the fields
$A_\mu ^{\alpha }, c^{\alpha  },\bar{c}^{\alpha }, B^{\alpha }$ are obtainable from the gauge fixed fermion, $\Psi^L$ as 
 \begin{eqnarray}
A^{\mu\alpha *} &= & \frac{\delta\Psi^L}{\delta A_\mu ^\alpha}= \partial^\mu
\bar{c}^\alpha \nonumber , \\
\bar{c}^{\alpha *} &= & \frac{\delta\Psi^L}{\delta\bar{c}^\alpha}=
[ \frac{\lambda}{2} B^\alpha - \partial\cdot A^\alpha ]\nonumber , \\
c^{\alpha *} & = & \frac{\delta\Psi^L}{\delta c^\alpha}=0\nonumber , \\
B^{\alpha *} & = & \frac{\delta\Psi^L}{\delta B^\alpha}=\frac{\lambda}{2}\bar{c}^\alpha.
\end{eqnarray}

Now,  we apply the FFBRST transformation given by Eq. (\ref{ffb}),
 to the above $Z^L$ where $\Theta(A,c,\bar{c},B)$ is chosen to have a particular form, 
obtainable from 
\begin{equation}
\Theta ^\prime (A,c,\bar{c},B)= i\int d^4 y\;\bar{c}^\alpha \left[\gamma_1\lambda B^\alpha +(\partial\cdot A^\alpha -\eta
\cdot A^ \alpha)\right], \label{ta}
\end{equation}
using Eq.(8). $\gamma_1$ is an arbitrary constant. We calculate the Jacobian by following the same procedure as discussed in 
section 3, which can be expressed as $e^{iS_1(\phi)}$ where
\begin{equation}
S_1= \int d^4 x\left[\gamma_1\lambda {(B^\alpha)}^2 + B^\alpha(\partial\cdot A^\alpha -\eta
\cdot A^ \alpha) -\bar{c}^\alpha (\eta^\mu - \partial^\mu) D^{\alpha\beta}_\mu c^\beta \right ].
\end{equation}
The transformed generating functional  becomes
\begin{eqnarray}
Z^{\prime L} &=& \int D\phi \exp\left\{ i\left [W^L_\Psi(\phi, \phi^*)+ S_1\right ]\right\},\nonumber\\
&=& \int D\phi \exp{\left[iW^A_\Psi(\phi, \phi^*)\right]}\equiv Z^A.
\end{eqnarray}
Under the FFBRST with parameter given in Eq. (\ref{ta}), $Z^L$ 
transforms to $Z^A$, which can be written explicitly in field/antifield formulation as
\begin{equation}
Z^A = \int [dAdcd\bar{c} dB] \exp \left[i\int d^4x \left\{ -\frac{1}{4}F^{\alpha\mu\nu}
F^\alpha_{\mu\nu}+\tilde{A}^{\mu\alpha *} D^{\alpha\beta}_\mu c^\beta+ \tilde{c}^{\alpha *}\frac{g}{2}
f^{\alpha\beta\gamma}c^\beta c^\gamma + B^\alpha \tilde{\bar{c}}^{\alpha *} 
\right\}\right].
\end{equation}
 In compact notation,
\begin{equation}
Z^A= \int D\phi \exp{\left [iS_0(\phi) 
+i \delta\Psi^A \frac{1}{\Lambda }\right ]},
\end{equation}
where
\begin{equation}
\Psi^A =\int d^4 x\; \bar{c}^\alpha\left [\frac{\xi}{2} B^\alpha - \eta\cdot A^\alpha \right ],
\end{equation}
with 
\begin{eqnarray}
\tilde{A}^{\mu\alpha *} &=& \frac{\delta\Psi^A}{\delta A_\mu ^\alpha}= -\bar{c}^\alpha\eta^\mu,
\nonumber \\
\tilde{\bar{c}}^{\alpha *} &=& \frac{\delta\Psi^A}{\delta\bar{c}^\alpha}=
\left[\frac{\xi}{2} B^\alpha - \eta\cdot A^\alpha \right],\nonumber \\
\tilde{c}^{\alpha *} &=& \frac{\delta\Psi^A}{\delta c^\alpha}=0,\nonumber \\
\tilde{B}^{\alpha *} &=& \frac{\delta\Psi^A}{\delta B^\alpha}=\frac{\xi}
{2}\bar{c}^\alpha.
\end{eqnarray}
$\xi=\;\lambda(1+2\gamma)$ is the gauge parameter in axial gauge.
 This $Z^A$ is the generating functional of YM theory for axial gauge in 
field/antifield formulation.
Thus, the FFBRST transformation given by Eq. (\ref{ffb}) with the parameter given in Eq. (\ref
{ta}) takes 
$Z^L $ to  $Z^A $. 
Both the extended actions in field/antifield formulation, $W^L_\Psi(\phi,\phi ^*)$
 and $W^A_\Psi(\phi,\tilde{\phi}^*)$ are the solutions of 
the master equation given in Eq. (\ref{ab}) and these are linked through FFBRST.\\
\\{\bf\small{Case II:\ Lorentz gauge to Coulomb gauge}}\\
\\Now, we consider another choice of FFBRST parameter $\Theta(A,c,\bar{c},B)$ corresponding to
\begin{equation}
\Theta^\prime (A,c,\bar{c},B)= i\int d^4 y \;\bar{c}^
\alpha \left[\gamma_1 \lambda B^\alpha + \partial_o A^\alpha_0 \right],\label{tc}
\end{equation}
and apply the FFBRST transformation  to $Z^L$. We calculate the Jacobian for this 
transformation which produces the additional term in the action 
\begin{equation}
S_1= \int d^4x \left[\gamma_1\lambda {(B^\alpha)}^2 + B^\alpha \partial_0A^\alpha_0 +\bar{c}^
\alpha{\partial ^0D_0}^{\alpha\beta}c^\beta\right].
\end{equation}
This yields
\begin{equation}
Z^{\prime L} = \int D\phi \exp\left\{i\left [W^L_\Psi(\phi, \phi^*)+ S_1\right ]\right\}\nonumber\\
\equiv Z^C,
\end{equation}
where 
\begin{equation}
Z^C = \int [dAdcd\bar{c} dB] \exp \left[i\int d^4x \left \{-\frac{1}{4}F^{\alpha\mu\nu}
F^\alpha_{\mu\nu}+\tilde{A}^{\mu\alpha *} D^{\alpha\beta}_\mu c^\beta+ \tilde{c}^{\alpha *}\frac{g}{2}
f^{\alpha\beta\gamma}c^\beta c^\gamma + B^\alpha \tilde{\bar{c}}^{\alpha *} 
\right\}\right ].
\end{equation}
Or, compactly
\begin{equation}
Z^C= \int D\phi \exp{\left [iS_0(\phi) 
+i \delta \Psi^C\frac{1}{\Lambda} \right ]}\equiv \int D\phi \exp{\left[iW^C_\Psi(\phi, \tilde{\phi}^*)\right]},
\end{equation}
where
\begin{equation}
\Psi^C=\int d^4 x \;\bar{c}^\alpha \left[\frac{\xi}{2}{B^\alpha} - \partial ^i A^{\alpha
}_i \right ],
\end{equation}
with \begin{eqnarray}
\tilde{A}^{{i\alpha}^*} &=& \frac{\delta\Psi^C}{\delta A_i ^\alpha}= 
-\bar{c}^\alpha \partial^i, \ \ \  \tilde{A}^{{0\alpha}^*}=0, \nonumber \\
\tilde{\bar{c}}^{\alpha *} &=& \frac{\delta\Psi^C}{\delta\bar{c}^\alpha}=\left[\frac{\xi
}{2}{B^\alpha} - \partial ^i A^{\alpha}_i \right ],\nonumber \\
\tilde{c}^{\alpha *} &=& \frac{\delta\Psi^C}{\delta c^\alpha}=0,\nonumber \\
\tilde{B}^{\alpha *} &=& \frac{\delta\Psi^C}{\delta B^\alpha}=\frac{\xi}{2}{\bar{c}^\alpha}
, \ \ \ \  i=1,2,3.
\end{eqnarray}
$\xi$ is the gauge parameter in Coulomb gauge. $Z^C$ is the generating functional for pure YM theory in Coulomb gauge in field/antifield formulation.
 Therefore, the FFBRST given by Eq. (\ref{ffb}) with parameter given in Eq. (\ref{tc}) connects the generating functional 
$ Z^L$ to $ Z^C$. Hence, the extended actions in Lorentz gauge $ 
W^L_\Psi(\phi,\phi^*)$
and in Coulomb gauge $ W^C_\Psi(\phi, \tilde{\phi}^*)$ which are the solutions of master 
equation are connected through FFBRST.\\
\\{\bf\small{Case III:\ Linear gauge to Quadratic gauge}}\\
\\Similarly, we choose another particular form of parameter $\Theta (A,c,\bar{c},B)$ related to
\begin{equation}
\Theta^\prime(A,c,\bar{c},B)= i \int d^4 y \; \bar{c}^\alpha \left[\gamma_1\lambda B^\alpha - d^{\alpha \beta \gamma }
A^\beta_\mu A^{\mu\gamma}\right], \label{tq}
\end{equation}
and perform FFBRST transformation to $Z^L$ as given in Eq. (\ref{zl}). The effective action 
$S_{eff}[\phi]$ in $Z^L$ will be modified due to the non-trivial Jacobian of these FFBRST 
transformation as $S_{eff}[\phi]$ + $S_1[\phi]$, where 
\begin{equation}
S_1 = \int d^4 x \left[\gamma_1\lambda {(B^\alpha)}^2-B^\alpha d^{\alpha \beta \gamma } A_\mu^\beta A ^{
\mu\gamma}-2 d^{\alpha \beta \gamma }\bar{c}^\alpha (D_\mu c)^\beta A^{\mu\gamma}\right]. \label{s1q}
\end{equation}
$d^{\alpha \beta \gamma} $ is a structure constant symmetric in $\beta$ and $\gamma$ 
. This 
additional term will transform the generating functional to 
\begin{equation}
Z^{\prime L} = \int D\phi \exp\left\{i\left [W^L_\Psi(\phi, \phi^*)+ S_1\right ]\right\}\nonumber\\
\equiv Z^Q.
\end{equation}
This $Z^Q$ is the generating functional for pure YM theory in quadratic gauge and is expressed  
in field/antifield formulation as 
\begin{equation}
Z^Q = \int [dAdcd\bar{c} dB] \exp \left[i\int d^4x \left\{-\frac{1}{4}F^{\alpha\mu\nu}
F^\alpha_{\mu\nu}+\tilde{A}^{\mu\alpha *} D^{\alpha\beta}_\mu c^\beta+ \tilde{c}^{\alpha *}\frac{g}{2}
f^{\alpha\beta\gamma}c^\beta c^\gamma + B^\alpha \tilde{\bar{c}}^{\alpha *} 
\right\}\right].
\end{equation}
Or, in compact notation
\begin{equation}
Z^Q= \int D\phi \exp{\left [iS_0(\phi) 
+i \delta \Psi^Q \frac{1}{\Lambda}\right ]}\equiv \int D\phi \exp{\left[iW^Q_\Psi(\phi, \tilde{\phi}^*)\right]},
\end{equation}
$W^Q_\Psi(\phi, \tilde{\phi}^*)$ is the extended action in quadratic gauge and
\begin{equation}
\Psi^Q=\int d^4 x \;\bar{c}^\alpha \left[\frac{\xi}{2}B^\alpha -(\partial \cdot {A^\alpha
} + d^{\alpha\beta\gamma} A_{\mu}^\beta A^{\gamma\mu})\right ],
\end{equation}
with \begin{eqnarray}
\tilde{A}^{{\mu\alpha}^*} &=& \frac{\delta\Psi^Q}{\delta A_\mu ^\alpha}= \partial^\mu
\bar{c}^\alpha - 2d^{\alpha\beta\gamma}\bar{c}^\beta A^{\mu\gamma}, \nonumber \\
\tilde{\bar{c}}^{\alpha *} &=& \frac{\delta\Psi^Q}{\delta\bar{c}^\alpha}=
\left[\frac{\xi}{2}B^\alpha -\left (\partial \cdot {A^\alpha
} + d^{\alpha\beta\gamma} A_{\mu}^\beta A^{\gamma\mu}\right )\right ],\nonumber \\
\tilde{c}^{\alpha *} &=& \frac{\delta\Psi^Q}{\delta c^\alpha}=0,\nonumber \\
\tilde{B}^{\alpha *} &=& \frac{\delta\Psi^Q}{\delta B^\alpha}=\frac{\xi}{2}\bar{c}^\alpha.
\end{eqnarray}
The FFBRST transformation with the parameter mentioned in Eq. (\ref{tq}), relates
$ Z^L$
to $Z^Q$. Hence, the solutions of master equation are linked through FFBRST.\\
\\{\bf\small{Case IV:\ Lorentz gauge (with parameter $\lambda$) to Lorentz gauge (with different parameter $\lambda^\prime$)}}\\
\\The discussion proceeds very much the same way as in earlier cases. We choose the FFBRST parameter
\begin{equation}
\Theta ^\prime (A,c,\bar{c},B)= i\gamma \int d^4 y\;\bar{c}^\alpha B^\alpha.
\end{equation}
The extra term in the action for this case is calculated as
\begin{equation}
S_1= \xi_1\int d^4x (B^\alpha)^2.
\end{equation}
The transformed generating functional is written in field/antifield formulation as
\begin{eqnarray}
Z^{L{(\lambda^\prime)}} &=& \int [dAdcd\bar{c} dB] \exp\left[i\int d^4x \left\{-\frac{1}{4}F^{
\alpha\mu\nu}F^\alpha_{\mu\nu}+\tilde{A}^{\mu\alpha *} D^{\alpha\beta}_\mu c^\beta+ \tilde
{c}^{\alpha *}\frac{g}{2}f^{\alpha\beta\gamma}c^\beta c^\gamma + B^\alpha \tilde{\bar{c}}^{
\alpha *} \right\}\right ],\nonumber \\
&=& \int D\phi\exp\left\{i\left[W^{L(\lambda)}_\Psi(\phi, 
\tilde{\phi}^*)+S_1\right]\right\}  \equiv \int D\phi\exp\left[iW^{L^{(\lambda^\prime)}}_\Psi (\phi, 
\tilde{\phi}^*)\right ].\nonumber
\end{eqnarray}
 $W^{L^{(\lambda^\prime)}}_\Psi(\phi, 
\tilde{\phi}^*)$ is the extended action in Lorentz gauge with gauge parameter $\lambda^\prime$
and
\begin{equation}
\Psi^{L^{(\lambda^\prime)}} =\int d^4 x \;\bar{c}^\alpha\left [\frac{\lambda^\prime}{2}B^\alpha -\partial\cdot A^\alpha \right ],
\end{equation}
with 
\begin{eqnarray}
\tilde{A}^{\mu\alpha *} &=& \frac{\delta\Psi^{L^{(\lambda^\prime)}}}{\delta A_\mu ^\alpha
}= \partial^\mu\bar{c}^\alpha, \nonumber \\
{\tilde{\bar{c}}}^{\alpha *} &=& \frac{\delta\Psi^{L^{(\lambda^\prime)}}}{\delta\bar{c}^
\alpha}=[\frac{\lambda^\prime}{2}B^\alpha-\partial\cdot A^\alpha ],\nonumber \\
\tilde{c}^{\alpha *} &=& \frac{\delta\Psi^{L^{(\lambda^\prime )}}}{\delta c^\alpha}=0,
\nonumber \\
\tilde{B}^{\alpha *} &=& \frac{\delta\Psi^{L^{(\lambda^\prime)}}}{\delta B^\alpha}=\frac{\lambda ^\prime}{2}\bar{c}^\alpha.
\end{eqnarray}
Both the extended actions in field/antifield formulation,
$ W^L_\Psi(\phi,\phi^*)$ and
$ W^{L^{(\lambda^\prime)}}_\Psi(\phi, \tilde{\phi}^*)$ 
are the solution of master equation given in Eq. (\ref{ab}) and are connected by FFBRST 
transformation.
\subsection{Connecting different solutions of master equation in effective theories}
In this section, we show that not only the solutions of master equation in the theory with different gauges 
are connected through FFBRST but the solutions of master equation in different effective 
theories are also linked through FFBRST. In particular, the solutions of master equation in BRST 
invariant theory is related to the solution of master equation in  through FFBRST. The 
generating functional for the most general BRST/anti-BRST invariant theory \cite{abrst} is 
written as 

\begin{equation}
Z^L_{AB} = \int [dAdcd\bar{c} dB] \exp{\left [ i S_{eff}^{AB}(A,c,\bar{c},B)\right ]},\label{zlab} 
\end{equation}
where
\begin{eqnarray}
S_{eff}^{AB}[A,c,\bar{c},B] &=& \int d^4x \left[-\frac{1}{4}F^{\alpha \mu \nu }F^{\alpha}_{\mu 
\nu} - \frac{(\partial\cdot A^\alpha)^2}{2\xi}  + \partial ^\mu\bar{c}D_\mu c\right. \nonumber \\
&+&\left.\frac{\alpha}{2} gf^{\alpha \beta \gamma }\partial\cdot A^\alpha\bar{c}^\beta c^\gamma -\frac{1}{8}\alpha (1-\frac{1}{2}\alpha )\xi g^2 f^{\alpha \beta \gamma}\bar{c}^\beta\bar{c}^\gamma f^{\alpha 
\eta \xi}c^\eta c^\xi \right ]. 
\end{eqnarray}
This effective action has the global symmetries under the following transformations.\\
BRST:
\begin{eqnarray}
\delta A_\mu ^\alpha &=&(D_\mu c)^\alpha \Lambda , \nonumber \\
\delta c^\alpha &=& -\frac{1}{2} g f^{\alpha \beta \gamma }c^\beta c^\gamma \Lambda , 
\nonumber \\
\delta\bar{c}^\alpha &=&\left(\frac{\partial\cdot A^\alpha}{\xi} -\frac{1}{2}\alpha g f^{
\alpha \beta \gamma} \bar{c}^\beta c^\gamma \right) \Lambda, \nonumber \\
\end{eqnarray}
anti-BRST:
\begin{eqnarray}
\delta A_\mu ^\alpha &=&(D_\mu \bar{c})^\alpha \Lambda , \nonumber \\
\delta c^\alpha &=& -\frac{1}{2} g f^{\alpha \beta \gamma }\bar{c}^\beta \bar{c}^\gamma 
\Lambda , 
\nonumber \\
\delta\bar{c}^\alpha &=&\left( -\frac{\partial\cdot A^\alpha}{\xi} -(1-\frac{1}{2}\alpha) g 
f^{\alpha \beta \gamma} \bar{c}^\beta c^\gamma \right) \Lambda . 
\end{eqnarray}\\
The most general BRST/anti-BRST effective action can be re-expressed in terms of auxiliary field $B^\alpha$ as 
\begin{eqnarray}
S_{eff}^{AB}[A,c,\bar{c},B] &=& \int d^4x \left[-\frac{1}{4}F^{\alpha \mu \nu }F^{\alpha}_{\mu 
\nu}+\frac{\xi}{2}(B^\alpha)^2 - B^\alpha(\partial\cdot A^\alpha-\frac{\alpha g \xi}{2}f^{\alpha\beta\gamma}\bar{c}^\beta c^\gamma)\right. \nonumber\\
&+&\left. \partial ^\mu\bar{c}D_\mu c -\frac{1}{8}\alpha\xi g^2f^{\alpha\beta\gamma}\bar{c}^\beta \bar{c}^\gamma f^{\alpha\eta\xi}c^\eta c^\xi \right].\label{seffab}
\end{eqnarray}
The off-shell nilpotent, global BRST/anti-BRST symmetries for the above effective action are given as\\
BRST:
\begin{eqnarray}
\delta A_\mu ^\alpha &=&(D_\mu c)^\alpha \Lambda , \nonumber \\
\delta c^\alpha &=& -\frac{1}{2} g f^{\alpha \beta \gamma }c^\beta c^\gamma \Lambda , 
\nonumber \\
\delta\bar{c}^\alpha &=& B^\alpha \Lambda,\nonumber\\
\delta B^\alpha&=& 0.
\end{eqnarray}
\\
anti-BRST:
\begin{eqnarray}
\delta A_\mu ^\alpha &=&(D_\mu \bar{c})^\alpha \Lambda , \nonumber \\
\delta \bar{c}^\alpha &=& -\frac{1}{2} g f^{\alpha \beta \gamma }\bar{c}^\beta \bar{c}^\gamma \Lambda , 
\nonumber \\
\delta c^\alpha &=& \left (-B^\alpha - g f^{\alpha \beta \gamma }\bar{c}^\beta c^\gamma \right ) \; \Lambda,\nonumber\\
\delta B^\alpha&=& -gf^{\alpha \beta \gamma }B^\beta \bar{c}^\gamma \; \Lambda.
\end{eqnarray}

Now, we choose a finite BRST parameter 
\begin{equation} 
\Theta^\prime (A,c,\bar{c},B)= i \int d^4 y\; \bar{c}^\alpha\left[\gamma_1\lambda B^\alpha -\frac{1}{2}\alpha g\xi f^{\alpha \beta \gamma } 
\bar{c}^\beta c^\gamma \right],\label{tab}
\end{equation}
and perform the FFBRST transformation given by Eq. (\ref{ffb}) to $Z^L$ written in Eq.
 (\ref{zl}). We calculate the Jacobian corresponding to this FFBRST transformation and expressed as $e^{iS_1(\phi)}$, where
\begin{equation}
S_1=\int d^4x \left[\gamma_1 \lambda {B^\alpha}^2+\frac{1}{2}\alpha g\xi f^{\alpha\beta\gamma} B^\alpha\bar{c}^\beta c^\gamma -\frac{1}{8}\alpha\xi g^2 f^{\alpha\beta\gamma}\bar{c}^\beta \bar{c}^\gamma f^{\alpha\eta \xi}c^\eta c^\xi \right ].
\end{equation}
This will give rise to a new generating functional 
\begin{equation}
Z^{\prime L} = \int D\phi \exp\left\{i\left [W^L_\Psi(\phi, \phi^*)+ S_1\right ]\right\},\nonumber\\
\equiv Z^L_{AB}.
\end{equation}
Under the choice of FFBRST parameter given in Eq. (\ref{tab}), $Z^L$ transforms to $Z^L_{AB}$ which can be expressed explicitly in field/antifield formulation as
\begin{equation}
Z^L_{AB} = \int [dAdcd\bar{c} dB] \exp \left[i\int d^4x \left \{-\frac{1}{4}F^{\alpha\mu\nu}
F^{\alpha}_{\mu\nu} +\tilde{A}^{\mu\alpha *} D^{\alpha\beta}_\mu c^\beta - \tilde{c}^{\alpha *}
\frac{g}{2}
f^{\alpha\beta\gamma}c^\beta c^\gamma + B^\alpha \tilde{\bar{c}}^{\alpha *} 
\right\}\right ].
\end{equation}
In compact notation,
\begin{equation}
Z^L_{AB}= \int D\phi \exp{\left [iS_0(\phi) 
+i \delta\Psi^L_{AB}\frac{1}{\Lambda}\right ]}\equiv \int D\phi \exp{\left[iW^L_{AB}(\phi, \tilde{\phi}^*)\right]},
\end{equation}
where
\begin{equation}
\Psi^L_{AB}=\int d^4 x \;\bar{c}^\alpha \left[\frac{\xi}{2}B^\alpha -\left (\partial \cdot A^\alpha - {\frac{\alpha g\xi} {4}} f^{\alpha \beta \gamma }\bar{c}^\beta c^\gamma \right )\right ],
\end{equation}
with 
\begin{eqnarray}
\tilde{A}^{\mu\alpha *} &=& \frac{\delta\Psi^L_{AB}}{\delta A_\mu ^\alpha}= \partial^\mu
\bar{c}^\alpha, \nonumber \\
\tilde{\bar{c}}^{\alpha *} &=& \frac{\delta\Psi^L_{AB}}{\delta\bar{c}^\alpha}=\left[\frac{\xi}{2}B^\alpha -\left (\partial \cdot A^\alpha - {\frac{\alpha g\xi} {2}} f^{\alpha \beta \gamma }\bar{c}^\beta c^\gamma\right ) \right ], \nonumber \\
\tilde{c}^{\alpha *} &=& \frac{\delta\Psi^L_{AB}}{\delta c^\alpha}= \frac{1}{4}\alpha g\xi  f^{\alpha \beta \gamma }\bar{c}^\beta \bar{c}^\gamma, \nonumber \\
\tilde{B}^{\alpha *} &=& \frac{\delta\Psi^L_{AB}}{\delta B^\alpha}=\frac{\xi}{2}\bar{c}^\alpha.
\end{eqnarray}
$\xi$ is the gauge parameter in the most general BRST/anti-BRST invariant theory. Therefore, the FFBRST given by Eq. (\ref{ffb}) with the parameter given in Eq. (\ref{tab}) takes $ Z^L= \int D\phi \exp \left [i 
W^L_\Psi(\phi,\phi^*) \right ]$ to $Z^L_{AB}= \int D\phi \exp \left [i 
W^L_{AB}(\phi,\tilde{\phi}^*) \right ].$

Both the extended action, $ W^L_\Psi(\phi,\phi^*)$ and
 $W^L_{AB}(\phi, \tilde{\phi}^*)$ in field/antifield formulation are the solution of master equation. 

\section{Conclusion}
In this paper, we explore the role of FFBRST transformation in field/antifield 
formulation. In field/antifield formulation the solutions of quantum master equation,  $W_\Psi[\phi,\phi^*]$ does not depend on the choice of $\Psi$, the gauge fixed fermion.We
have considered FFBRST transformation in the field/antifield formulation of YM 
theories and have shown that it connects the generating functional, corresponding to different 
solutions of master equation by considering  several explicit  examples. Particularly, we have shown that 
the generating functional $Z^L$ in Lorentz gauge corresponding to the solution of master 
equation, $W^L_\Psi[\phi,\phi^*]$ is connected to (i) generating functional $Z^A$ corresponding 
to the solution of master equation in axial gauge $W^A_\Psi[\phi,\phi^*]$. (ii) generating 
functional $Z^C$ corresponding to the solution of master equation in Coulomb gauge,
$W^C_\Psi[\phi,\phi^*]$ .(iii) generating 
functional $Z^Q$ corresponding to the solution of master equation in quadratic gauge,
$W^Q_\Psi[\phi,\phi^*]$ and, (iv) generating 
functional $Z^{L{(\lambda^\prime)}}$ corresponding to the solution of master equation in 
different gauge parameter,
 $W^{L{(\lambda^\prime)}}_\Psi[\phi,\phi^*]$.
 FFBRST not only connects theories with different solutions of master equations in different 
gauges but also it connects different $W_\Psi[\phi,\phi^*]$ corresponding to different 
effective theories. $W^L_\Psi[\phi,\phi^*]$ in BRST invariant theory is connected to 
$W^L_{AB}[\phi,\phi^*]$, corresponding to the most general BRST/anti-BRST invariant theory 
through FFBRST. In all the cases the non-trivial Jacobians of FFBRST play the crucial role.\\

\noindent
{\Large{\bf {Acknowledgment}}} \\

\noindent
We thankfully acknowledge the financial support
from the Department of Science and Technology (DST), Government of India, under
the SERC project sanction grant No. SR/S2/HEP-29/2007.\\

\end{document}